\documentstyle[12pt]{article}
\def\beq{\begin{equation}}
\def\eeq{\end{equation}}

\begin{document}

\begin{titlepage}
\begin{center}
{\Large \bf Theoretical Physics Institute \\
University of Minnesota \\}  \end{center}
\vspace{0.3in}
\begin{flushright}
TPI-MINN-92/61-T \\
November 1992
\end{flushright}
\vspace{0.4in}
\begin{center}
{\Large \bf Some properties of amplitudes at multi boson thresholds in
spontaneously broken scalar theory\\}
\vspace{0.2in}
{\bf M.B. Voloshin  \\ }
Theoretical Physics Institute, University of Minnesota \\
Minneapolis, MN 55455 \\
and \\
Institute of Theoretical and Experimental Physics  \\
Moscow, 117259 \\
\vspace{0.2in}
{\bf   Abstract  \\ }
\end{center}

It is shown that in a $\lambda \phi^4$ theory of one real scalar field with
spontaneous breaking of symmetry a calculation of the amplitudes of
production by a virtual field $\phi$ of $n$ on-mass-shell bosons all being
exactly at rest is equivalent in any order of the loop expansion to a
Euclidean space calculation of the mean field of a kink-type configuration.
Using this equivalence it is found that all the $1 \to n$ amplitudes have no
absorptive part at the thresholds to any order of perturbation theory. This
implies non-trivial relations between multi-boson threshold production
amplitudes.  In particular the on-mass-shell amplitude of the process $2 \to
3$ should vanish at the threshold in all loops. It is also shown that the
factor $n!$ in the $1 \to n$ amplitudes at the threshold is not eliminated
by loop effects.

\end{titlepage}

The tree-level amplitudes of the process in which one virtual scalar field
$\phi$ produces $n$ on-mass-shell bosons of this field in a $\lambda \phi^4$
theory displays a growth proportional to $n!$ at large $n
{}~^{\cite{cornwall}-\cite{akp1}}$. If this growth is not eliminated by the
loop effects this would imply that a strong interaction develops in a theory
with weak coupling $\lambda$ at energies where production of $n > 1/\lambda$
interacting bosons is kinematically possible. It was recently found that in
a $\lambda \phi^4$ theory of a scalar field $\phi$ one can explicitly sum
all tree graphs for the process $1 \to n$ at the threshold, i.e. when all
the produced particles have zero spatial momenta, both in the theory without
spontaneous symmetry breaking$^{\cite{v1}}$ and in the case of the
spontaneous breaking of the reflection symmetry $\phi \to -\phi
{}~^{\cite{akp1}}$. An elegant development of these calculations was
subsequently suggested by Brown$^{\cite{brown}}$ who had pointed out that
the threshold amplitudes of the $1 \to n$ processes are related to a complex
vacuum solution of the field equations. An extension of Brown's technique
has enabled explicit calculation of these amplitudes at the one-loop
level$^{\cite{v2,smith1,smith2}}$. In particular it was found that the
relative magnitude of the loop corrections is described by the parameter
$n^2 \lambda$ so that the quantum effects may indeed completely modify the
behavior of the amplitudes at $n \ge \lambda$. A surprising result of this
study was that the on-mass-shell scattering amplitudes for the processes
$2 \to n$ at the tree level are vanishing at the $n$ particle threshold for
all $n$ greater than 4 in unbroken $\lambda \phi^4$ theory$^{\cite{v2}}$ and
for $n > 2$ in the theory with spontaneously broken
symmetry$^{\cite{smith1}}$.  One of the consequences of the latter behavior
is that the threshold amplitudes for the $1 \to n$ processes do not develop
absorptive part at the one-loop level in the theory with the spontaneous
symmetry breaking.

In this paper we consider the threshold production of particles in the
theory of one real scalar field with spontaneously broken symmetry beyond
the one-loop approximation. This is possible due to the fact that the loop
expansion around the Brown's solution can be mapped into the loop expansion
in the Euclidean space around a kink-type configuration of the field. As a
result it will be shown that the threshold amplitudes of the $1 \to n$
processes have no absorptive part to any order in this expansion. Through
the unitarity relation this implies that the on-mass-shell amplitude of the
process $2 \to 3$ should vanish at the threshold in all orders of
perturbation theory, while for higher final state multiplicities in the
on-mass-shell processes $k \to n$ this implies existence of non-trivial
relations between amplitudes with different $k$. Also we will use the
analytical properties of the time dependence of the Brown type solution to
prove that the generating function for the amplitudes of $1 \to n$ has a
singularity at a finite distance in the complex plane, so that the $n!$
growth survives all the loop effects.

The theory to be discussed in this paper is described in the Minkowski space
by the Lagrangian density

\beq
L = {1 \over 2} (\partial_\mu \phi)^2 + {{m^2} \over 4} \phi^2 - {\lambda
\over 4} \phi^4~.
\label{lagrang}
\eeq
The field has two degenerate vacuum states, which in the classical
approximation correspond to $\phi = \pm v$ with $v=m/\sqrt{2 \lambda}$ and
$m$ is the mass of bosons propagating in either of the two vacua. For
definiteness we will discuss scattering processes in the ``left" vacuum
$\phi=-v$. The amplitudes of the process $1 \to n$ are determined by the
matrix elements

\beq
a_n = \langle n | \phi(0) | 0 \rangle~,
\label{me}
\eeq
where $\langle n |$ is the final state of the $n$ bosons, which at the
threshold all have exactly zero spatial momenta, ${\bf p}_a=0$. Using the
standard reduction formula Brown$^{\cite{brown}}$ has noticed that the sum

\beq
\Phi(t)=\sum_{n=0}^\infty {{z_0^n \, e^{i n m t}} \over {n!}} \langle n |
\phi(0) | 0 \rangle = \sum_{n=0}^\infty {{z_0^n} \over {n!}} \langle n |
\phi(t) | 0 \rangle
\label{bsum}
\eeq
is given by the vacuum-to-vacuum matrix element of the field $\phi(t)$

\beq
\Phi(t)=\langle 0_{\rm out} | \phi(t) | 0_{\rm in} \rangle^\rho
\label{vtv}
\eeq
in the presence of the source $\rho(t)$ in the following limit. The source
is  chosen in the form $\rho(t)=\rho_\omega \exp (i \omega t)$ and then
$\omega $ is taken to the mass shell, $\omega \to m$, simultaneously with
tending $\rho_\omega$ to zero in such a way that the quantity

\beq
z(t)={{\rho_\omega e^{i \omega t}} \over {m^2-\omega^2-i\epsilon}}
\label{zt}
\eeq
tends to a finite limit $z(t) = z_0 e^{i m t}$. Since the source drives only
the positive-frequency response of the field and the amplitude of the source
is vanishing on the mass shell, this limiting procedure can be replaced by
the requirement that $\Phi(t)$ is a solution of the field equations which
admits expansion in ascending powers of $z(t)$, i.e. it has only the
positive frequency part.

The solution of the classical Euler-Lagrange equation

\beq
{{d^2} \over {dt^2}} \Phi - {{m^2} \over 2} \Phi + \lambda \Phi^3 =0~,
\label{class}
\eeq
subject to this condition, thus generates the expressions for the matrix
elements (\ref{me}) as given by the sum of tree graphs. The explicit form of
the classical solution is

\beq
\Phi_{cl}(t)=-v{{1+z(t)/(2 v)} \over {1-z(t)/(2 v)}}
\label{bsol}
\eeq
and thus the matrix elements (\ref{me}) are given by

\beq
\langle n | \phi(0) | 0 \rangle= \left ( {\partial } \over {\partial z}
\right )^n \Phi |_{z=0} = -n! (2v)^{1-n}, ~~~~~~n \ge 1
\label{tres}
\eeq
which reproduces the previously known result$^{\cite{akp1}}$.

The loop corrections in this approach arise from quantum corrections to the
response of the field to the source in the described above limit. In the one
loop approximation these corrections in the theory under discussion were
calculated by Smith$^{\cite{smith1}}$:

\beq
\Phi_{one-loop}(t)=-{\bar v}{{1+z(t)/(2 {\bar v})} \over {1-z(t)/(2
{\bar v})}} - v {{\sqrt{3}\,\lambda} \over {2 \pi}} {{[z(t)/(2v)]^2}
\over {[1-z(t)/(2v)]^3}} ~,
\label{oneloop}
\eeq
where ${\bar v}$ is an appropriately renormalized vacuum expectation
value. From this result one easily finds the amplitudes with the one-loop
correction$^{\cite{smith1}}$:

\beq
\langle n | \phi(0) | 0 \rangle= -n! (2{\bar v})^{1-n} \left (1 + n (n-1)
{{\sqrt{3}\,\lambda} \over {8 \pi}} \right )~.
\label{l1me}
\eeq

One can readily notice that the explicit expressions as well as the initial
definition (\ref{bsum}) define $\Phi(t)$ in terms of a Taylor series in the
variable $z(t)$ with a finite radius of convergence. The quantity $z_0$ in
the equation (\ref{bsum}) in fact serves as a ``tag" of the variable $z(t)$
and can be chosen arbitrarily small, but finite, within the radius of
convergence of the series. It can be also noticed that the problem of
whether the $n!$ growth of the amplitudes survives the loop effects is
equivalent to the problem of whether the latter effects keep the radius of
convergence of the series finite, or whether they suppress the amplitudes to
the extent that this radius becomes infinite, in which case the $n!$ growth
disappears.  This is one of the central questions to be addressed in this
paper. Another essential point is that the expansion is indeed of the Taylor
series type.  This behavior is ensured by the time evolution of the
Heisenberg operator $\phi(t)$:

\beq
\langle n | \phi(t) | 0 \rangle = e^{i E_n t} \langle n | \phi(0) | 0 \rangle
\label{evol}
\eeq
where $E_n$ is the energy of the state $\langle n |$ relative to the vacuum.
In the standard setting of the scattering problem the energy of $n$
particles all being at rest in the asymptotic $out$ state is always the sum
of their energies, so that the expansion for $\Phi(t)$ is polynomial in
$\exp (i m t)$ in a field theory. In the case of Quantum Mechanics the
behavior is different:  high levels in an anharmonic oscillator are not
equidistant, hence the present analysis can not be directly applied in that
case. This helps to resolve the conflict between the absence of the
factorial growth of the amplitudes in Quantum Mechanics and the possibility
of such growth in the field theory. We will return to discussion of this
point in the concluding part of this paper.

If $z_0$ is chosen within the radius of convergence of the Taylor series,
the generating function $\Phi(t)$ does not have singularities in the
upper half-plane of complex $t$, where the Taylor series does converge. In
the lower half-plane the classical solution (\ref{bsol}) as well as the one
loop correction (\ref{oneloop}) develops singularities, which are located
with the period $2 \pi /m$ parallel to the real axis of $t$ (see Fig. 1).
Higher loops produce mounting singularities at these points, near which the
perturbation theory breaks down.

For the following we
introduce the variable $\tau$ in such way that $z(t)/(2 v)=-e^{m \tau}$, so
that

\beq
\tau=i t+ i {\pi \over m} + {1 \over m} \ln {{z_0} \over {2 v}}~.
\label{tau}
\eeq
In terms of the variable $\tau$ the classical solution has the familiar
simple form of the kink:

\beq
\Phi_{cl}(\tau)= v \, \tanh {{m \tau} \over 2}~.
\label{tanh}
\eeq
The calculation of the first quantum correction to the classical field
$\Phi$ amounts to the calculation of the tadpole graph of Fig.2, where the
propagator is the Green function in the background $\Phi_{cl}$. In the
course of this calculation it was found$^{\cite{v2,smith1}}$ that the
appropriate Green function satisfies the condition that it vanishes, when
either of its arguments corresponds to either $\tau \to -\infty$ or $\tau \to
+\infty$.

The convergence of the Taylor series for $\Phi(t)$ in the upper half-plane
of $t$ implies that the series is convergent for sufficiently large in
absolute value negative $\tau$. Therefore one can impose the boundary
condition on the solution $\Phi(t)$ to the full quantum problem at $\tau \to
-\infty$ by requiring that the asymptotic behavior there is of the form

\beq
\Phi(t) \to -v + v e^{m\tau}, ~~~~~\tau \to -\infty~.
\label{bc}
\eeq

Now we have the following ingredients, which relate our problem to that of
calculating the full quantum mean field of the kink configuration: the
zeroth order approximation (equation (\ref{tanh})), the condition on the
Green function, which ensures that the quantum corrections vanish at both
infinities in $\tau$, and, finally, the boundary condition (\ref{bc}), which
fixes the normalization of the first term in the expansion in powers of
$e^{m \tau}$ and thus fixes the translational zero mode in the kink
background.  These ingredients are sufficient to relate the problem of
calculation of $\Phi(\tau)$ to that of calculating the mean field in the
Euclidean-space field theory, i.e. the path integral

\beq
\Phi(\tau)= {{\int \phi(x) \, e^{-S[\phi]}\, {\cal D}\phi} \over
\int  \, e^{-S[\phi]}\, {\cal D}\phi }
\label{pint}
\eeq
with

\beq
S[\phi]= \int \left ({1 \over 2} (\partial_\mu \phi)^2 - {{m^2} \over 4}
\phi^2 + {\lambda \over 4} \phi^4~ \right ) \, d^4 x
\label{action}
\eeq
and with the kink-type boundary conditions, i.e. the condition in
the equation (\ref{bc}) and also $\phi(\tau \to +\infty)=v$, the latter is
ensured by the asymptotic behavior at $\tau \to +\infty$ of the classical
solution (\ref{tanh}) and the condition that the Green function is vanishing
at $\tau \to +\infty$.

Under the specified boundary conditions the path integral in the equation
(\ref{pint}) is well defined. Indeed, the classical solution (\ref{tanh})
provides the absolute minimum of the action (\ref{action}) under these
boundary conditions. The four dimensional theory is free from infra-red
divergencies and the ultra-violet ones are as usual compensated by the
counterterms for the mass and the coupling constant renormalization. Thus we
arrive at the conclusion that the mean field $\Phi(\tau)$ defined by the
integral (\ref{pint}) is real and the same is true for the coefficients
$c_n$ of its expansion in ascending powers of $e^{m \tau}$ at $\tau \to
-\infty$:

\beq
\Phi(\tau)=\sum_{n=0}^\infty c_n e^{n m \tau}~.
\label{tauexp}
\eeq
Therefore we conclude that the matrix elements (\ref{me})

\beq
\langle n | \phi(0) | 0 \rangle = (-1)^n n! c_n
\label{mec}
\eeq
are real in all orders of perturbation theory. In the one-loop approximation
this non-trivial behavior is seen from the explicit result (equation
(\ref{l1me})), and we see that this behavior extends to all orders of the
loop expansion.

Before proceeding to discussion of the implications of this result, we can
mention an example of a theory where such behavior does not hold. This is
the theory with unbroken symmetry, i.e. with positive mass term in the
Lagrangian. In that theory the classical solution $\Phi_{cl}(\tau)$ provides
a saddle point of the action, rather than the minimum, hence the integration
over the negative mode yields the mean field in the Euclidean space
problem essentially complex already in the one-loop
approximation$^{\cite{v2}}$.

The absence of an absorptive part of the amplitudes $a_n$ imposes
through the unitarity relation non-trivial conditions on the on-shell
amplitudes. At the one-loop level the condition is that the on-mass-shell
scattering amplitudes for the processes $2 \to n$ vanish in the tree
approximation for all $n$ greater than 2. This result follows from the
fact that the only possible cuts of the one-loop graphs, are those across
two lines, and those cuts split the graphs into tree diagrams. In higher
loops unitary cuts across more than two lines are possible, so that the
imaginary part can be zero as a result of cancellation between contributions
of intermediate states with different number of particles. So in general one
would find relations between the amplitudes of the processes $k \to n$ with
different $k$. There is however one special case in which one can find a
well formulated conclusion. Namely, let us consider the amplitude of the
process $1 \to 3$. According to the previous discussion its imaginary part
is zero to all orders in perturbation theory. On the other hand, the only
intermediate state which could give rise to an imaginary part is
that with two particles, i.e. 2\,Im$A(1 \to 3)=A(1 \to 2) \cdot A(2 \to 3)$.
For three intermediate particles there is no phase space (recall that the
final three bosons are produced at rest), while for larger number $k$ of
particles in the intermediate state the on-mass-shell process $k \to 3$ is
kinematically impossible, when the final particles are at rest.
Furthermore for the scattering $2 \to 3$ at the threshold there is no
angular dependence, i.e. only the $S$ wave can be present, therefore no
cancellation between the partial waves can occur. Therefore, since the
amplitude $A(1 \to 2)$ is manifestly non zero, one concludes that the
on-mass-shell amplitude $A(2 \to 3)$ is vanishing at the threshold to all
orders of the loop expansion.

We now proceed to the most intriguing question of whether the factorial
behavior of the matrix elements (\ref{mec}) at large $n$ holds in all loops.
This is certainly true if the Taylor expansion (\ref{tauexp}) in the
variable $u=e^{m \tau}$ has a finite radius of convergence $R$. Then the
coefficients $c_n$ behave at large $n$ as $|c_n|=\zeta (n) \, R^{-n}, $
where $\zeta(n)$ behaves weaker than exponent of $n$ at large $n$, and thus
the behavior of the coefficients $c_n$ would not compensate the $n!$ in
eq.(\ref{mec}). We will assume the opposite, i.e. that the series in
eq.(\ref{tauexp}) has infinite radius of convergence and show that this
assumption is inconsistent.

If the Taylor series (\ref{tauexp}) for the mean field $\Phi(\tau)$ has
infinite radius of convergence, the same is true for the quantity

\beq
\Delta(\tau)=v^2-\Phi^2(\tau)
\label{diff}
\eeq
(notice that it is the square of the mean field entering here and not the
mean value of the square of the field). The asymptotic behavior of
$\Delta(\tau)$ at $\tau \to -\infty$ is $\Delta(\tau) \to 2v^2 e^{m \tau}$.
At $\tau \to + \infty$ the asymptotic behavior is determined by the
boundary condition on $\Phi(\tau)$ and by the fact that $m$ is the lowest
energy in the spectrum: $\Delta(\tau) \to b e^{-m \tau}$ where $b$ is a
constant.

The series (\ref{tauexp}) defines $\Phi(\tau)$ and thus also $\Delta(\tau)$
as a periodic function of complex $\tau$ with the period $2 \pi /m$. Let us
consider the contour ${\cal C}$ in the complex plane of $\tau$ shown in Fig.
3. Two links of the contour run along the lines separated by one period:
Im$\, \tau=0$ and Im$\, \tau = 2 \pi/m$, and the horizontal links between
the two lines are chosen sufficiently far at negative and positive Re$\,
\tau$, so that one can use the asymptotic expression for $\Delta(\tau)$ on
those links. One can now consider calculation of the index of the function
$\Delta(\tau)$ on the contour ${\cal C}$:

\beq
I = {1 \over {2 \, \pi \, i}} \oint_{\cal C} {{d \Delta(\tau) /d \tau} \over
{\Delta(\tau)}} \, d\tau~,
\label{ind}
\eeq
which gives the difference between the number of  zeros and poles of the
function $\Delta(\tau)$ inside the contour: $I=Z - P$, where $Z$ is the
number of zeros and $P$ is the number of poles inside the contour. Since the
function $\Delta(\tau)$ is assumed to have no singularities at finite
complex $\tau$, the only obstacle to calculating the index integral
(\ref{ind}) can be presence of isolated zeros of $\Delta(\tau)$ at the
contour ${\cal C}$. This obstacle however can be
easily removed by slightly shifting the contour away from each of the
isolated zeros.  (Clearly, a condensation of zeros of $\Delta(\tau)$ at a
finite $\tau$ would either contradict the assumption that $\Delta(\tau)$ is
entire function, or imply that the function is identically zero, which
obviously is not the case.) One can readily see, that the boundary condition
that $\Delta(\tau)$ falls exponentially at both infinities in Re$\, \tau$
completely determines the index (\ref{ind}) :  $I=-2$. Therefore
$\Delta(\tau)$ necessarily has a non-zero number of poles inside the contour
${\cal C}$, which is inconsistent with the assumption that the series
(\ref{tauexp}) has infinite radius of convergence. This completes the proof
that the amplitudes (\ref{mec}) have a factorial growth in $n$ in the full
quantum theory.

In the context of the present discussion one can illustrate the fundamental
difference between the field theory matrix elements $\langle n | \phi(0) | 0
\rangle$, which grow proportionally to $n!$ and seemingly analogous matrix
elements $\langle n | x | 0 \rangle$ in Quantum Mechanics of an anharmonic
oscillator, described by the Hamiltonian

\beq
H = {{p^2} \over 2} + {{x^2} \over 2} + {\lambda \over 4} x^4~,
\label{qm}
\eeq
where the frequency of the oscillator is set to one for simplicity. The
matrix element of $x$ between the ground state $|0 \rangle$ and the $n$-th
excited state $| n \rangle$ can be evaluated by the Landau WKB
method$^{\cite{LWKB}}$ and has different behavior in the two regions: $n
\lambda \ll 1$, $\langle n | x | 0 \rangle \sim n! \lambda^{n/2}$ and
$n\lambda \gg 1$, $\langle n | x | 0 \rangle \sim e^{-n}$, and in both cases
$n \gg 1$ is assumed (see e.g.  in Ref.\cite{v3}).  Naturally, in the
Quantum Mechanics of one degree of freedom there is no possibility for an
unlimited growth with $n$ of these matrix elements. This can be
seen$^{\cite{brown}}$ e.g.  from the sum rule

\beq
\langle 0 |x^2| 0 \rangle = \sum_n |\langle n | x | 0 \rangle|^2~,
\label{sr}
\eeq
so that the sum on the right hand side has to be finite.
The crucial difference from the field theory is that the analog of the sum
(\ref{bsum})

\beq
X(t) = \sum_{n=0}^\infty {{z_0^n} \over {n!}} \langle n | x(t) | 0 \rangle =
\sum_{n=0}^\infty {{z_0^n} \over {n!}} e^{i E_n t} \langle n | x(0) | 0
\rangle
\label{qmsum}
\eeq
for the Heisenberg operator $x(t)$ is not an expansion in integer powers of
$e^{i t}$, since the energies $E_n$ of the anharmonic oscillator are not
equidistant, when $n\lambda > 1$.  Therefore one can not apply to $X(t)$ the
Abel's theorems about power series expansion of an analytical function and
the singularities of the function $X(t)$ in the complex plane are not
related to the asymptotic behavior of the matrix elements $\langle n | x | 0
\rangle$ in the same simple way as the asymptotic growth of the matrix
elements (\ref{me}) is related to the existence of the singularities of
$\Phi(t)$ at finite complex $t$.

Thus it has been shown that the problem of calculating the amplitude of the
process $1 \to n$ at the threshold in the $\lambda \phi^4$ theory of one
real scalar field $\phi$ is equivalent to calculating in the corresponding
Euclidean field theory the mean field $\Phi(\tau)$ (equation (\ref{pint}))
with the kink-type boundary conditions. The fact that the classical solution
provides the absolute minimum of the action in the Euclidean-space problem
leads to the conclusion that all the $1 \to n$ amplitudes at the threshold
have vanishing absorptive part despite the presence in higher loops of many
intermediate states with on-mass-shell particles. This behavior in
particular requires the on-mass-shell amplitude of the scattering $2 \to 3$
to vanish at the threshold in all orders of perturbation theory. Also using
the periodicity of the mean field $\Phi(\tau)$ in complex $\tau$ and the
behavior of the field at large $| {\rm Re}\, \tau|$ it has been shown that
the function $\Phi(\tau)$ necessarily has singularities at finite complex
$\tau$ and thus the threshold amplitudes $1 \to n$, which are related to the
coefficients of the expansion of $\Phi(\tau)$ in powers of $e^{m \tau}$
(eqs.(\ref{tauexp}) and (\ref{mec})), necessarily have a factorial growth in
$n$ in the full quantum theory.  The properties of the specific theory,
which were employed in the present paper, in general do not hold
for other models. In the quantum-mechanical example of an anharmonic
oscillator the spectrum of energies of highly excited states $|n \rangle$ is
not equally spaced, thus the generating function can not be treated in terms
of a power series expansion in $e^{i t}$, neither it is periodic in $t$. In
the $\lambda \phi^4$ theory without spontaneous symmetry breaking or in a
theory with the spontaneous breaking of a continuous symmetry the
Euclidean-space classical solution for the analog of $\Phi(\tau)$ does not
correspond to the minimum of the Euclidean action, so that a proper
definition of the Euclidean-space path integral in the full quantum theory
is somewhat subtle.  It can be believed however, that the latter subtlety
can eventually be sorted out and that the amplitudes in those theories also
can be shown to display the $n!$ growth in all orders of perturbation
theory, unlike the case of Quantum Mechanics. However this is yet to be
done.

I am thankful to Larry McLerran and Valery Rubakov for numerous discussions
and their critical remarks. This work is supported in
part by the DOE grant DE-AC02-83ER40105.

{\Large \bf Figure captions}\\[0.3in]
{\bf Figure 1.} The structure of the mean field $\Phi(t)$ in the complex $t$
plane. $\Phi(t)$ is analytical in the upper half-plane. The heavy dots
correspond to the poles of the classical mean field $\Phi_{cl}(t)$. The
vertical line, running between two poles is the real axis for
$\tau$.\\[0.15in]
{\bf Figure 2.} The tadpole graph for calculating the one-loop correction to
the mean field (the heavy dot). The propagator and the triple vertex are the
Green function and the vertex in the background of the classical mean
field $\Phi_{cl}(t)$.\\[0.15in]
{\bf Figure 3.} The contour ${\cal C}$ on which the index of the function
$\Delta(\tau)$ is calculated (equation (\ref{ind})).

\newpage
\unitlength=1mm
\thicklines
\begin{picture}(121.00,105.00)
\put(40.00,70.00){\line(1,0){80.00}}
\put(80.00,105.00){\line(0,-1){70.00}}
\put(70.00,105.00){\vector(0,-1){70.00}}
\put(80.00,50.00){\circle*{2.00}}
\put(100.00,50.00){\circle*{2.00}}
\put(60.00,50.00){\circle*{2.00}}
\put(40.00,50.00){\circle*{2.00}}
\put(120.00,50.00){\circle*{2.00}}
\put(117.00,71.00){\makebox(0,0)[cb]{{\Large Re$\,t$}}}
\put(81.00,103.00){\makebox(0,0)[lc]{{\Large Im$\, t$}}}
\put(69.00,37.00){\makebox(0,0)[rc]{{\Large Re$\, \tau$}}}
\put(80.00,15.00){\makebox(0,0)[cc]{{\Large \bf Figure 1}}}
\end{picture}

\vspace{1.5 in}

\begin{picture}(98.00,37.00)
\put(60.00,30.00){\circle*{2.00}}
\put(91.00,30.00){\circle{14.00}}
\put(60.00,30.00){\line(1,0){24.00}}
\put(80.00,00.00){\makebox(0,0)[cc]{{\Large \bf Figure 2}}}
\end{picture}

\newpage

\begin{picture}(120.00,120.00)
\put(40.00,70.00){\line(1,0){80.00}}
\put(70.00,120.00){\line(0,-1){100.00}}
\put(70.00,30.00){\vector(0,-1){10.00}}
\put(70.00,30.00){\line(1,0){20.00}}
\put(90.00,30.00){\line(0,1){80.00}}
\put(90.00,110.00){\line(-1,0){20.00}}
\put(90.00,60.00){\vector(0,1){20.00}}
\put(115.00,71.00){\makebox(0,0)[cb]{{\Large Im$\, \tau$}}}
\put(72.00,22.00){\makebox(0,0)[lc]{{\Large Re$\, \tau$}}}
\put(89.00,85.00){\makebox(0,0)[rc]{{\Large ${\cal C}$}}}
\put(69.00,69.00){\makebox(0,0)[rt]{{\Large 0}}}
\put(91.00,69.00){\makebox(0,0)[lt]{{\Large ${{2 \pi} \over m}$}}}
\put(80.00,5.00){\makebox(0,0)[cc]{{\Large \bf Figure 3}}}
\end{picture}

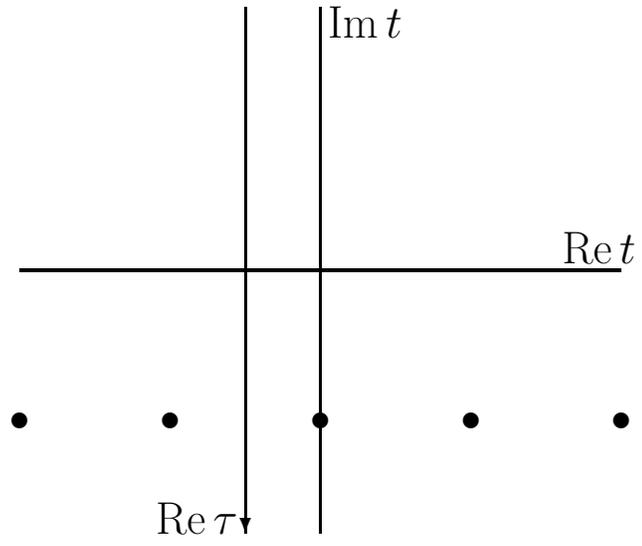
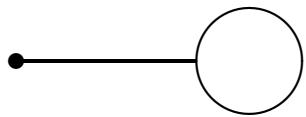
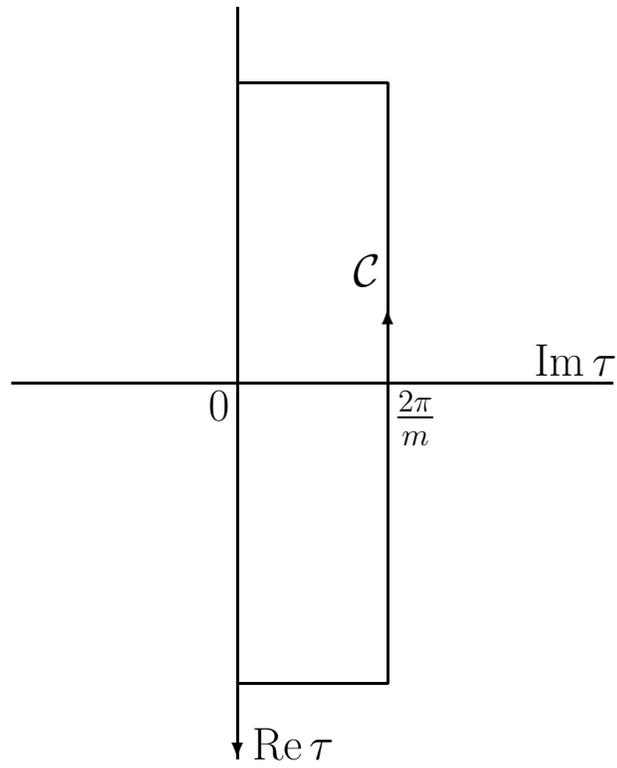
\end{document}